\documentclass[12pt,a4paper]{article}

\pdfoutput=1
\usepackage{amsmath}
\usepackage{amsfonts}
\usepackage{amssymb}
\usepackage{graphicx}
\usepackage{array}
\usepackage{cancel}
\usepackage{caption}
\usepackage{wrapfig}
\usepackage{mathrsfs}
\usepackage{cancel}
\usepackage{siunitx}
\usepackage{amsthm}
\usepackage{tikz-cd}
\usetikzlibrary{babel}

\usepackage[left=1.5cm,right=1.5cm,top=1.5cm,bottom=1.5cm,includefoot,footskip=1.5cm]{geometry}
\usepackage[OT1, T2A]{fontenc}
\usepackage[utf8]{inputenc}
\usepackage[english]{babel}
\usepackage[toc]{appendix}
\usepackage{csquotes}

\usepackage[sorting=none, citestyle=numeric-comp,
backend=biber
]{biblatex}
\usepackage{biblatex2bibitem}

\usepackage[colorlinks=true,
linkcolor=blue,citecolor=green,breaklinks=true]{hyperref}

\let\{=\lbrace
\let\}=\rbrace
\let\[=\lbracket
\let\]=\rbracket

\let\w = \omega

\newcommand\ili{\int\limits}
\newcommand {\lb} {\left(}
\newcommand {\rb} {\right)}

\newcommand\p{\partial}

\newcommand{\pdf}[2]{\dfrac{\partial #1}{\partial #2}}

\newcommand{\bl}{\left\langle}
\newcommand{\br}{\right\rangle}

\newcommand{\mt}[1]{\mathcal{#1}}

\DeclareMathOperator{\res}{res}

\numberwithin{equation}{section}
\usepackage{authblk}

\author[1,2,3]{I.V.Kochergin}
\affil[1]{Moscow Institute of Physics and Technology, Institutskii per. 9, 141700, Dolgoprudny, Russia}
\affil[2]{Landau Institute for Theoretical Physics, 142432, Chernogolovka, Russia}
\affil[3]{Skolkovo Institute of Science and Technology, 121205, Moscow, Russia}

\title{On Calabi-Yau manifolds in weighted projective spaces and their mirror GLSMs}

\begin{document}

\maketitle

\begin{abstract}

The goal of the present paper is to calculate the complex structure moduli space K\"ahler potentials for hypersurfaces in weighted projective spaces and compare with the partition functions of their mirror GLSMs. We explicitly perform the K\"ahler potential computation and show that the corresponding formula is well-defined in case of quasismooth hypersurfaces. We then construct the mirror GLSM with an appropriate number of K\"ahler parameters and discuss the interpretation of its partition function in terms of mirror symmetry. Namely, it is shown that different contributions to the partition function are related to various charts of the complex structure moduli space.

\end{abstract}
\newpage

\section{Introduction}\label{intro}
The special K\"ahler geometry of Calabi-Yau moduli spaces naturally emerges in the context of superstring theory compactifications: it determines the Yukawa couplings of the effective low-energy theory \cite{CANDELAS_Moduli1991, Strominger:1990}. Due to this connection, we will consider the case of threefolds in this work. The moduli space $M[X]$ of a given CY manifold $X$ is a direct product of two distinct parts $M_C[X]$ and $M_K[X]$, associated with the deformations of complex and K\"ahler structures of $X$ respectively. Their dimensions can be expressed in terms of Hodge numbers: $\dim M_C[X] = h^{2,1}(X)$, $\dim M_K[X] = h^{1,1}(X)$. In string theory, $M_K[X]$ is naturally complexified, and its geometry acquires instanton corrections \cite{DINE1986}, while $M_C[X]$ (which is classically complex) stays the same on the quantum level. Thus, the geometry of $M_C[X]$ can be computed purely from geometric considerations. In fact, the K\"ahler potential $K_C(X)$ on $M_C[X]$ can be expressed via the $(3,0)$-holomorphic form $\Omega$:
\begin{equation}
e^{-K_C(X)} = \ili_X \Omega \wedge \bar{\Omega}.
\end{equation}
The explicit use of this formula which involves the computation of periods of $\Omega$ over cycles in $H_3(X)$, is rather complicated and can be performed explicitly only in a limited number of cases \cite{CANDELAS1991, Candelas1994}. However, in the works \cite{AB, Aleshkin_2018, Aleshkin:2017fuz, Aleshkin:2018jql} a new method was developed, which is based on the connection between $H_3(X)$ and a relative homology group $H_5^+(\mathbb{C}^5)$. It works when the family of manifolds in $M_C$ is defined as zero locus of a homogenous polynomial in a weighted projective space. The K\"ahler potential is then expressed as a power series in deformation parameters~--- in the case of a large complex structure regime, an analytic continuation is required. We will discuss this approach in greater detail in section \ref{geomc}.

The quantum-corrected K\"ahler potential on $M_K[X]$, which we denote $K_K(X)$, can be recovered via mirror symmetry \cite{CANDELAS1991}. Namely, there is a map between CY manifolds which interchanges the moduli spaces: if $Y$ is a mirror to $X$, then $M_K[X] = M_C[Y]$, $M_C[X] = M_K[Y]$. We will use two particular mirror manifold construnctions: of Batyrev \cite{Batyrev1993DualPA, ARTEBANI2016319} and of Berglund, Hubsch and Krawitz (BHK) \cite{Berglund:1991pp, krawitz2009fjrw}.

Yet another way to compute $K_K(X)$ was conjectured in \cite{J}. In this paper $K_K(X)$ is connected with a partition function $Z_{\text{GLSM}}(X)$ of a Gauged Linear Sigma Model (GLSM) whose vacuum moduli space is $X$:
\begin{equation}
e^{-K_K(X)} = Z_{\text{GLSM}}(X).
\end{equation}
The explicit formula for $Z_{\text{GLSM}}$ was found in \cite{Doroud_2013, Benini_2014}, also see \cite{GomisExact} for the physical proof of the conjecture. We will be mainly interested in the mirror symmetry version of this formula. As we show in section \ref{partfnc}, Batyrev's construction of a mirror manifold naturally yields the corresponding sigma model. Hence, if $Y$ is a mirror to $X$, we have:
\begin{equation}
e^{-K_C(X)} = Z_{\text{GLSM}}(Y).
\end{equation}
In principle, it gives a different method to compute $K_C(X)$. The result obtained this way works for the entire moduli space, unlike the geometric one. However, we will see that they coincide only if the GLSM has a particular Landau-Ginzburg phase, while in other cases the additional contributions to the partition function can be attributed to different charts of the moduli space $M_C[X]$.

The rest of the paper is organized as follows. In section \ref{geomc} we define the class of CY manifolds we work with and their complex structure moduli spaces. We then briefly discuss the K\"ahler potential computation method and obtain an explicit formula under some additional assumptions. Section \ref{partfnc} is dedicated to the construction of the mirror GLSM and the computation of its partition function. We also compare the result with the K\"ahler potential and discuss the source of the discrepancy. Besides, we investigate the possible mirror symmetry interpretation. Some additional technical details and particular examples are provided in appendices. 

\section{Geometric computation}\label{geomc}
\subsection{Hypersurfaces in weighted projective spaces}
Following \cite{AB} we consider the hypersurface defined by zeros of a quasi-homogeneous polynomial $W(x)$ in a weighted projective space $\mathbb{P}^4_{w_1,\dots,w_5}$: $W(\lambda^{w_1}x_1,\dots,\lambda^{w_5}x_5) = \lambda^w W(x)$, where $w = w_1 + \dots + w_5$. From geometric point of view $W(x)$ is a section of the anticanonical bundle $\mt{O}(w)$, so its zero locus defines CY manifold (not necessarily smooth) \cite{Hori:2003ic}. Moreover, different polynomials correspond to hypersurfaces which are isomorphic as real, but generally not as complex manifolds. Hence they can be regarded as different points in complex structure moduli space. In order to do the practical calculations we need to choose the <<reference point>> $W_0(x)$:
\begin{equation}
W_0(x) = \sum_{i=1}^5 \prod_{j=1}^5 x_j^{M_{ij}}\label{W0}
\end{equation}
We assume that $M$ is invertible and $W_0$ is \emph{transverse}, i.e. the only solution of $\p_i W_0(x) = 0$ is $x=0$. The latter condition means that the hypersurface $X_0: W_0(x) = 0$ is quasismooth, i.e. all of its singularities are those of the weighted projective space \cite{AB}. All such polynomials were classified in \cite{Kreuzer_1992}, see also \cite{Aleshkin:2019ahf} for more practical notation.

Two polynomials describe the same manifold if they are related via a coordinate transformation. Therefore we can fix the following form of $W(x)$:
\begin{equation}
W(x) = W_0(x) + \sum_{a=1}^h \phi_a e_a,\quad e_a = \prod_{j=1}^5 x_j^{S_{aj}},\label{W(x)}
\end{equation}
where $\{e_a\}$ span a basis in the degree $w$ part of the Milnor ring $R(W_0)$:
\begin{equation}
R(W_0) = \dfrac{\mathbb{C}[x_1,\dots,x_5]}{\bl \p_1 W_0,\dots,\p_5 W_0\br},
\end{equation}
and $\phi_a$ are complex deformation parameters. Its convenient description for transverse polynomials is presented in \cite{Aleshkin:2019ahf}. Note that in generic situation $h \le h^{2,1}$ (see \cite{Candelas:1994bu} for example), so in this way we can describe only some subspace of the actual moduli space. Hence our computations will in fact be restricted to this subspace.

We also can slightly generalize this construction by taking an additional quotient by a finite group $G$ with diagonal action (i.e. $x_i \mapsto \w_i x_i$ for $\w = (\w_1,\dots,\w_5) \in  G$). This action should preserve $W_0(x)$ and the holomorphic form $\Omega_0$ on $W_0(x) = 0$, which can be expressed as follows \cite{AB}:
\begin{equation}
\Omega_0 = \text{Res}_{W_0(x) = 0} \dfrac{x_5 dx_1\wedge\dots \wedge dx_4}{W_0(x)}.
\end{equation}
Thus we find that the monomial $x_1 x_2 x_3 x_4 x_5$ should be invariant. Note that it is also of degree $w$, and in most cases, it belongs to $R(W_0)$ (if, for instance, the highest degrees of $x_i$ in $W_0$ are greater than $2$, we will consider only such cases). So there is always at least one deformation of the form $e_1 = x_1x_2x_3x_4x_5$. Of course, after taking the quotient by $G$, all other deformations should be $G$--invariant as well. 

It is also convenient to consider the transposed polynomial
\begin{equation}
 W_0^T(x) = \sum_{i=1}^5\prod_{j=1}^5 x_j^{M_{ji}},\label{trpol}
\end{equation}
i.e. the one given by $M^T$ instead of $M$. It is quasi-homogeneous with respect to a different set of weights $\{\tilde{w}_i\}$ and has a degree $\tilde{w} = \sum_{i=1}^5 \tilde{w}_i$ (we assume that they are positive integers with minimal possible values). Its zero locus defines a hypersurface in $\mathbb{P}_{\tilde{w}_1,\dots,\tilde{w}_5}^4$ which (or its quotient by a finite group) is in fact the mirror of $X_0$ according to the BHK construction \cite{Berglund:1991pp}. We will return to this fact later in section \ref{partfnc}, for now we will use $\tilde{w}_i$ to simplify some formulas.  
 
\subsection{K\"ahler potential formula}
The method of K\"ahler potential computation in \cite{AB} is based on using specific relative cohomology of $\mathbb{C}^5$. Namely, we define the following differential:
\begin{equation}
D_- = d - d W_0 \wedge,\quad D_-^2 = 0,
\end{equation}
and consider the respective cohomology group $H^5_{D_-}(\mathbb{C}^5)$ which is isomorphic to $R(W_0)$. It is dual to the relative homology group $H_5^+(\mathbb{C}^5) = H_5(\mathbb{C}^5,~ \text{Re}\, W_0 \to +\infty)$ (so we allow the cycles to have endpoints at infinity with $\text{Re}\, W_0 = +\infty$), the corresponding pairing is as follows:
\begin{equation}
\bl \mt{Q},f \br = \ili_{\mt{Q}} f e^{-W_0},~ f \in H^5_{D_-}(\mathbb{C}^5),~\mt{Q} \in H_5^+(\mathbb{C}^5).\label{pair}
\end{equation}
It is straightforward to see that the integral converges and vanishes for $D_-$--exact forms. 

Next we consider the subgroup $\mt{H}^5$ of $H_{D_-}^5(\mathbb{C}^5)$ invariant under $\mt{G} = \mathbb{Z}_w$ which acts as follows:
\begin{equation}
x_i \mapsto \w^{w_i} x_i,\quad \w = \exp\lb \frac{2\pi i k}{w}\rb \in \mathbb{Z}_w,~k\in\mathbb{Z}
\end{equation}
In case of an additional quotient by $G$ we should use $\mt{G} = \mathbb{Z}_w \oplus G$. The convenient basis is $\{e_\alpha d^5 x\}$, where $\{e_\alpha\}$ are $\mt{G}$--invariant monomials in the basis of $R(W_0)$, $\alpha = 0,\dots,2h+1$. It has a natural grading corresponding to the degree of monomials (under $x_i \to \lambda^{w_i} x_i$) divided by $w$. The only element of degree $0$ is $e_0 = 1$, there is also only one element of degree $3$: $e_{2h+1} \sim \det (\p_i \p_j W_0)$. It is however more convenient to use the following representation:
\begin{equation}
e_{2h+1} = \prod_{i,j=1}^5 x_j^{M_{ij}-2}.\label{e2h1}  
\end{equation}
The degree $1$ elements are essentially $e_a$ and $e_{h+a} = \frac{e_{2h+1}}{e_a}$ are elements of degree $2$. One can also define the following pairing:
\begin{equation}
\eta(f,g) = \res_{x = 0} \dfrac{f g\, d^5 x}{\prod_{i=1}^5 \p_i W_0},~f\,d^5x,g\,d^5x~\in \mt{H}^5.\label{pairet}
\end{equation}
In our basis it is antidiagonal: $\eta_{\alpha\beta} = \eta(e_\alpha,e_\beta) \sim \delta_{\alpha, 2h+1-\beta}$. In fact, there is an injective map $\mt{H}^5 \to H^3(X_0:~W_0(x) = 0)$ such that the grading corresponds to Hodge decomposition and $\eta$ to Poincare pairing (up to a sign). Of course, it is an isomorphism when $h = h^{2,1}$ \cite{Aleshkin_2018}.

The homology group $\mt{H}_5$ dual to $\mt{H}^5$ can be defined as a quotient of $H^+_5(\mathbb{C}^5)$ by its subgroup orthogonal to $\mt{H}^5$. Then it is convenient to define the dual basis of cycles $\{\Gamma_\alpha\}$:
\begin{equation}
\ili_{\Gamma_\alpha} e_\beta e^{-W_0} d^5 x = \delta_{\alpha\beta}\label{Galpha}
\end{equation}
These cycles are not geometric: for instance, if $\{\mt{Q}_\alpha\}$ is some basis of geometric cycles, we find:
\begin{equation}
\mt{Q}_\alpha = T_{\alpha\beta} \Gamma_\beta,\quad T_{\alpha\beta} = \ili_{\mt{Q}_\alpha} e_\beta e^{-W_0} d^5x.\label{Qc}
\end{equation}
As we will see in the next subsection, matrix elements $T_{\alpha\beta}$ are essentially complex. Also, the fact that $\mt{Q}_\alpha$ are geometric means that they are real, so the complex--conjugated cycles are as follows:
\begin{equation}
\bar{\Gamma}_\alpha = \bar{T}^{-1}_{\alpha\gamma} T_{\gamma\beta} \Gamma_{\beta}.
\end{equation}
It means that $\mt{M}_{\alpha\beta} = T^{-1}_{\alpha\beta} \bar{T}_{\beta\gamma}$ is a real structure matrix, also it does not depend on the choice of the real cycles $\mt{Q}_\alpha$.

The main result of \cite{AB, Aleshkin_2018, Aleshkin:2017fuz, Aleshkin:2018jql} is the following formula:
\begin{equation}
e^{-K_C(X)} = \sum_{\alpha,\beta,\gamma = 0}^{2h+1} (-1)^{|\gamma|} \sigma_\alpha(\phi) \eta_{\alpha\beta}\mt{M}_{\beta\gamma} \bar{\sigma}_\gamma(\bar{\phi}).\label{KPot1}
\end{equation}
Here $|\alpha|$ is a degree of $e_\alpha$ divided by $w$, $X$ is a zero locus of (\ref{W(x)}) in $\mathbb{P}^4_{w_1,\dots,w_5}$ and $\sigma_\alpha$ are periods over cycles $\Gamma_\alpha$:
\begin{equation}
\sigma_\alpha = \ili_{\Gamma_\alpha} e^{-W} d^5 x.\label{persig}
\end{equation}
As the exponential K\"ahler potential itself is defined up to multiplication by (locally) holomorphic and antiholomorphic functions, we can forget about the constant in $\eta_{\alpha\beta}$ and consider $\eta_{\alpha\beta} = \delta_{
\alpha,2h+1-\beta}$. This formula was used in papers \cite{AB,Aleshkin_2018, Aleshkin:2017fuz, Artemev:2020gaq, Belakovskiy:2020nno} to compute $e^{-K_C(X)}$ in cases of particular polynomials. Besides, in \cite{Aleshkin:2018jql} it was computed for all \emph{Fermat} polynomials (i.e. when $M$ is diagonal). The general formula was conjectured in \cite{Aleshkin:2019ahf}, and in the rest of this section we will derive it and show that it is well-defined.
\subsection{Real structure matrix}
First, let us find the real structure matrix $\mt{M}_{\alpha\beta}$. To simplify the expression, we define the following matrices:
\begin{equation}
B = M^{-1},\quad \hat{S}:~ e_\alpha = \prod_{j=1}^5 x_j^{\hat{S}_{\alpha j}}
\end{equation}
In particular, $\hat{S}$ has the following properties:
\begin{equation}
\hat{S}_{\alpha j} = \hat{S}_{2h+1,j} - \hat{S}_{2h+1-\alpha,j},\quad\hat{S}_{0j} = 0,\quad\hat{S}_{a j} = S_{aj},~a=1,\dots,h.\label{prS}
\end{equation}
There are also some properties attributed to the fact that $W$ is quasi-homogeneous:
\begin{equation}
\sum_{j=1}^5 M_{ij} w_j = \sum_{j=1}^5 S_{aj}w_j = {w},\quad \sum_{j=1}^5 B_{ij} = \dfrac{w_i}{w},\quad \sum_{i=1}^5 B_{ij} = \dfrac{\tilde{w}_j}{\tilde{w}}.\label{tilw}
\end{equation}
Besides, due to (\ref{e2h1}) we have:
\begin{equation}
\hat{S}_{2h+1,j} = \sum_{i=1}^5 M_{ij} - 2,\quad \hat{S}_{2h+1,k} B_{kj} = 1 - 2\dfrac{\tilde{w}_j}{\tilde{w}}.\label{prSB}
\end{equation}

In order to define the cycles $\{\mt{Q}_\alpha\}$, we make a change of coordinates $x_i = \prod_{j=1}^5 y_j^{B_{ij}}$. Its Jacobian is as follows:
\begin{equation}
J(y) = \det \pdf{x_i}{y_j} = \det B \prod_{i=1}^5 \dfrac{x_i}{y_i} = \det B \prod_{j=1}^5 y_j^{{\tilde{w}_j}/{\tilde{w}} - 1}.
\end{equation}
As $W_0(y) = \sum_{i=1}^5 y_i$ it is natural to assume that the cycle $\mt{Q}_\alpha$ can be factorized into a product of one-dimensional cycles: $\mt{Q}_\alpha = \prod_{i=1}^5 \mt{Q}_{\alpha}^i$, where $\mt{Q}_\alpha^i$ belongs to the complex plane of $y_i$. We also have a requirement that $\text{Re}\,W_0 \to +\infty$ at infinity, so the convenient choice of $\mt{Q}_\alpha^i$ is as follows: it goes from $+\infty$ to $0$ along the real line, turns around $0$ couterclockwise $N_{\alpha i}$ times ($N_{\alpha i} \in \mathbb{Z}$) and goes back to $+\infty$. This follows the construction proposed in \cite{Aleshkin:2019pmz}. The respective integral can be easily computed in terms of gamma functions:
\begin{equation}
\begin{aligned}
T_{\alpha\beta} &= \ili_{\mt{Q}_\alpha} e_\beta e^{-W_0} d^5 x = \det B\ili_{\mt{Q}_\alpha} \prod_{j=1}^5 \lb y_j^{\hat{S}_{\beta k} B_{kj} + \tilde{w}_j/\tilde{w}-1} e^{-y_j}\rb d^5 y =\\
&=\det B \prod_{j=1}^5 \left[ \lb e^{2\pi i N_{\alpha j} (\hat{S}_{\beta k} B_{kj} + \tilde{w}_j/\tilde{w})} - 1 \rb \Gamma\lb \hat{S}_{\beta k} B_{kj} + \tilde{w}_j/\tilde{w} \rb \right].\label{Tab}
\end{aligned}
\end{equation}
Note that this expression is well-defined even at the poles of the gamma function, i.e., when the expression $S_{\beta k} B_{kj} + \tilde{w}_j/\tilde{w}$ is a negative integer since it has a finite limit due to the prefactor. 

To proceed with calculations we need to find the relation between $T$ and $\bar{T}$. We have:
\begin{equation}
\begin{aligned}
    \bar{T}_{\alpha\beta} &= \det B\prod_{j=1}^5 \left[ \lb e^{2\pi i N_{\alpha j} (1-\hat{S}_{\beta k} B_{kj} - \tilde{w}_j/\tilde{w})} - 1 \rb \Gamma\lb \hat{S}_{\beta k} B_{kj} + \tilde{w}_j/\tilde{w} \rb \right] = \\
    &=\det B \prod_{j=1}^5 \left[ \lb e^{2\pi i N_{\alpha j} (\hat{S}_{2h+1-\beta, k} B_{kj} + \tilde{w}_j/\tilde{w})} - 1 \rb  \Gamma \lb \hat{S}_{2h+1-\beta, k} B_{kj} + \tilde{w}_j/\tilde{w} \rb \gamma \lb \hat{S}_{\beta k} B_{kj} + \tilde{w}_j/\tilde{w} \rb \right] = \\
    &= T_{\alpha,2h+1-\beta} \prod_{j=1}^5 \gamma \lb \hat{S}_{\beta k} B_{kj} + \tilde{w}_j/\tilde{w} \rb ,
\end{aligned}\label{Tbarab}
\end{equation}
where $\gamma(x) = \frac{\Gamma(x)}{\Gamma(1-x)}$. Here we used the properties (\ref{prS}) and (\ref{prSB}).  Hence, as $\bar{T}_{\alpha\beta} = T_{\alpha\gamma}\mt{M}_{\gamma\beta}$, the real structure matrix is as follows:
\begin{equation}
\mt{M}_{\alpha\beta} = \delta_{\alpha,2h+1-\beta} \prod_{j=1}^5 \gamma \lb \hat{S}_{\beta k} B_{kj} + \tilde{w}_j/\tilde{w} \rb.\label{rstruct}
\end{equation}
Some of the matrix elements might be singular if there is such $b$ that $S_{b k} B_{kj} + \tilde{w}_j/\tilde{w} = m \in \mathbb{Z}$. However, as we prove in appendix \ref{A}, transverse polynomials have the following important property: if $\{n_j\}$ is a set of non-negative integers, then the number of positive integer elements of $\{n_k B_{kj} + \tilde{w}_j /\tilde{w}\}$ is always not less than the number of negative integer elements. In the rest of this paper we will refer to it as property \ref{A}. It means that in fact there are no singularities~--- the pole of the numerator is always of not greater order than the one of the denominator.

Still, if there are such $b$ that $S_{bk}B_{kj} + \tilde{w}_j/\tilde{w} \in \mathbb{Z}$, and hence
\begin{equation}
\hat{S}_{2h+1-b,k}B_{kj} + \tilde{w}_j/\tilde{w} = 1 -(\hat{S}_{b k} B_{kj} + \tilde{w}_j/\tilde{w}) \in \mathbb{Z},\label{connform}
\end{equation}
we have $T_{\alpha b} = T_{\alpha,2h+1-b} = 0$ due to the same property \ref{A}. Therefore, in such cases $T$ is degenerate, meaning that our choice of $\mt{Q}_\alpha$ is bad. On the other hand, we can still compute $e^{-K_C(X)}$ using (\ref{rstruct}), and we will show that it is well-defined and, in some sense, coincides with the result of GLSM computation. Formally it can be explained in terms of the following regularization procedure. We deform $W_0$ so that the integration measure acquires an additional factor:
\begin{equation}
    W_0 \to W_0 - i\sum_{j} p_j  \log \prod_{k=1}^5 x_k^{M_{jk}},\quad d^5 y\,e^{-W_0} \to d^5 y\,\prod_{j=1}^5 y_j^{ip_j} e^{-W_0},\quad p_j \in \mathbb{R},\label{shift}
\end{equation}
hence the integration over $\mt{Q}_\alpha$ is still well-defined. Also note that the symmetries of $W_0$ given by diagonal action of $x_j$ are preserved. The new measure simply amounts to the shift $\tilde{w}_j/\tilde{w} \to \tilde{w}_j/\tilde{w} + ip_j$ in all formulas, so (\ref{connform}) cannot be satisfied and $T_{\alpha\beta}(p)$ is non-degenerate. By simple calculation similar to (\ref{Tbarab}) we find:
\begin{equation}
    \mt{M}_{\alpha\beta}(p) = \delta_{\alpha,2h+1-\beta} \prod_{j=1}^5 \gamma(\hat{S}_{\beta k} B_{kj} + \tilde{w}_j/\tilde{w} + i p_j).
\end{equation}
We see that after taking the limit $p_j \to 0$ it coincides with $\mt{M}_{\alpha\beta}$ from (\ref{rstruct}).

\subsection{Periods and the final answer}\label{sec24}
In order to calculate the periods $\sigma_\alpha$, we will use the recursive technique proposed in \cite{AB}. First, we expand the exponential of the deformations:
\begin{equation}
e^{-W} = e^{-W_0} \prod_{a=1}^h \sum_{n_a=0}^{\infty} (-1)^{n_a}\dfrac{\phi_a^{n_a}}{n_a!} e_a^{n_a} \label{eW}
\end{equation}
Then according to (\ref{persig}) we need to calculate the integrals of $\prod_{a=1}^h e_a^{n_a} d^5x$ over $\Gamma_\alpha$ with respect to the pairing (\ref{pair}). Hence, we can treat $\prod_{a=1}^h e_a^{n_a} d^5x$ as a cohomology class in $\mt{H}^5$~--- up to a $D_-$-exact form it should be some linear combination of $\{e_\beta d^5x\}$. Knowing the coefficients one can then evaluate the integral using (\ref{Galpha}). The reduction to $\{e_\beta d^5x\}$ is most easily performed step by step. For instance one has the following cohomologic equivalence relations \cite{Aleshkin:2019pmz}:
\begin{equation}
\prod_{j=1}^5 x_j^{b_j} d^5 x \sim (b_i B_{ij} + \tilde{w}_j/\tilde{w}-1) \prod_{k=1}^5 x_k^{b_k - M_{jk}} d^5x.\label{cohomrel}
\end{equation}
Note that if the LHS is invariant with respect to some group $G$ which preserves $W_0$, the RHS is also invariant. These relations allow to consequently reduce the degree of $\prod_{j=1}^5 x_j^{b_j}$ and arrive to one of the $\{e_\alpha\}$ monomials and as we just discussed such procedure is consistent with taking an additional quotient. So, if $b_j = \hat{S}_{\beta j} + l_i M_{ij}$, $l_i \in \mathbb{Z}$ for some $\beta$ we find:
\begin{equation}
\mt{I}_\alpha(b) = \ili_{\Gamma_\alpha} \prod_{j=1}^5 x_j^{b_j} e^{-W_0} d^5x = \delta_{\alpha\beta} \prod_{j=1}^5 \dfrac{\Gamma(b_k B_{kj} + \tilde{w}_j/\tilde{w})}{\Gamma(\hat{S}_{\beta k} B_{kj} + \tilde{w}_j/\tilde{w})}.\label{Ial}
\end{equation}
One can check this directly by calculating the integral over $\mt{Q}_\alpha$ and using (\ref{Qc}) with the explicit expression for $T_{\alpha\beta}$:
\begin{equation}
\ili_{\mt{Q}_\alpha} \prod_{j=1}^5 x_j^{b_j} e^{-W_0} d^5 x = \det B \prod_{j=1}^5 \left[ \lb e^{2\pi i N_{\alpha j} (\hat{S}_{\beta k} B_{kj} + \tilde{w}_j/\tilde{w})} - 1 \rb \Gamma\lb b_k B_{kj} + \tilde{w}_j/\tilde{w} \rb \right] = T_{\alpha\beta}\mt{I}_\beta(b).\label{ilQ}
\end{equation}

The gamma functions in the numerator of (\ref{Ial}) might be singular. Suppose that $b_k B_{kj} + \tilde{w}_{j}/\tilde{w} \in \mathbb{Z}$ for exactly $l$ values of $j$, then the number of singular gamma functions is not greater than $l/2$ due to the property \ref{A}. Besides, $\hat{S}_{\beta k}B_{kj} + \tilde{w}_j/\tilde{w} \in \mathbb{Z}$ in such cases due to the expression for $b_j$. We can use (\ref{connform}):
\begin{equation}
    \hat{S}_{\beta k}B_{kj} + \tilde{w}_j/\tilde{w} = 1 - (\hat{S}_{2h+1-\beta, k}B_{kj} + \tilde{w}_j/\tilde{w}).\label{SBeq}
\end{equation}
Then by applying the property \ref{A} to $\hat{S}_{2h+1-\beta, k}B_{kj} + \tilde{w}_j/\tilde{w}$ we find that there are at least $l/2$ singular gamma functions in the denominator. Therefore (\ref{Ial}) is finite. Also by using \ref{A} for both sides of (\ref{SBeq}) one can show that $l$ should be even and precisely $l/2$ elements of $\{\hat{S}_{2h+1-\beta, k}B_{kj} + \tilde{w}_j/\tilde{w}\}$ are positive integers. Note that if $b_k B_{kj} + \tilde{w}_j/\tilde{w} \in \mathbb{Z}$ for some $j$, then $\mt{J}_\alpha(b) = 0$ despite that $\prod_{j=1}^5 x_j^{b_j}d^5x$ corresponds to the non-trivial element $e_\beta d^5 x \in \mt{H}^5$. Once again it reflects the fact that our choice of $\mt{Q}_\alpha$ is unsuitable for such cases, so the calculations should be understood in the sense of the limit discussed at the end of the previous subsection. We will describe it in more details later.

It is also important to understand when $e(b)d^5x = \prod_{j=1}^5 x_j^{b_j} d^5 x$ is non-trivial in $\mt{H}^5$. It means that one of the monomials from the basis, say $e_\alpha$, can be replaced with $e(b)$ and the corresponding new set of dual cycles $\{\Gamma_\beta\}$ is well-defined, i.e. $\mt{M}$ is non-degenerate and non-singular. Besides, we should also be able to replace $e_{2h+1-\alpha}$ with $e(\bar{b})$, $\bar{b}_j = \hat{S}_{2h+1,j} - b_j$, as it has a non-trivial pairing (\ref{pairet}) with $e(b)$: 
\begin{equation}
    \eta(e(b),e(\bar{b})) = 1,
\end{equation}
here we omitted the constant in $\eta$. Elements of $\bar{b}$ might be negative, but in the previous subsection we discussed that the integrals over $\mt{Q}_\beta$ are well-defined regardless. We also assume that we do not replace $e_1$ or $e_{2h+1}$. Via the calculations similar to (\ref{Tab}) and (\ref{Tbarab}) used for the new basis of monomials we find:
\begin{equation}
    \mt{M}_{\alpha, 2h+1-\alpha} = \prod_{j=1}^5 \gamma(b_k B_{kj} + \tilde{w}_j/\tilde{w}).
\end{equation}
We will denote this expression by $\mt{M}(b)$. Due to the property \ref{A} it is finite if $b_j \ge 0$. Hence, we assume that $e(b) d^5 x$ is non-trivial iff $\mt{M}(b) \neq 0$. Using \ref{A}, this condition can be reformulated as follows: the number of integer elemets of $\{b_kB_{kj} + \tilde{w}_j/\tilde{w}\}$ is even and precisely half of them are positive. Before we have shown that it is satisfied for the original basis of monomials.

Now we have all of the ingredients to write the final answer. Using (\ref{KPot1}), (\ref{rstruct}), (\ref{eW}) and (\ref{Ial}) and omitting the constant factor in $\eta_{\alpha\beta}$ we find:
\begin{equation}
\begin{aligned}
e^{-K_C(X)} &= \sum_{\alpha=0}^{2h+1} (-1)^{|\alpha|} |\sigma_\alpha(\phi)|^2 \prod_{j=1}^5\gamma(\hat{S}_{\alpha k}B_{kj} + \tilde{w}_j/\tilde{w}),\\
\sigma_{\alpha}(\phi) &= \sum_{\substack{n_a=0\\ (n_a S_{ak} - \hat{S}_{\alpha k})B_{kj} \in \mathbb{Z}}}^{\infty} \prod_{j=1}^5 \dfrac{\Gamma(n_a S_{ak}B_{kj} + \tilde{w_j}/\tilde{w})}{\Gamma(\hat{S}_{\alpha k}B_{kj} + \tilde{w}_j/\tilde{w})} \prod_{a=1}^h \dfrac{(-1)^{n_a} \phi_a^{n_a}}{n_a!}.\label{KalPotF}
\end{aligned}
\end{equation}
This expression coincides with one of \cite{Aleshkin:2019ahf}. 
Let us also slightly transform it. First, note that
\begin{equation}
\begin{aligned}
\Sigma_\alpha(n) &=    (-1)^{|\alpha|}\prod_{j=1}^5 \dfrac{\Gamma(n_a S_{ak}B_{kj} + \tilde{w_j}/\tilde{w})}{\Gamma(\hat{S}_{\alpha k}B_{kj} + \tilde{w}_j/\tilde{w})}\prod_{a=1}^h(-1)^{n_a} =\\
&=(-1)^{|\alpha|} \prod_{j=1}^5 \dfrac{\Gamma(n_a S_{ak}B_{kj} + \tilde{w_j}/\tilde{w})\Gamma(\hat{S}_{\alpha k} B_{kj} + \tilde{w}_j/\tilde{w})}{\gamma(\hat{S}_{\alpha k} B_{kj} + \tilde{w}_j/\tilde{w})\Gamma(\hat{S}_{\alpha k} B_{kj} + \tilde{w}_j/\tilde{w}) \Gamma(1-\hat{S}_{\alpha k} B_{kj}- \tilde{w}_j/\tilde{w})}\prod_{a=1}^h(-1)^{n_a} =\\
    &=\prod_{j=1}^5\dfrac{\sin[\pi (n_a S_{ak}B_{kj} + \tilde{w}_j/\tilde{w})]\Gamma(n_a S_{ak}B_{kj} + \tilde{w_j}/\tilde{w})\Gamma(\hat{S}_{\alpha k} B_{kj} + \tilde{w}_j/\tilde{w})}{\pi\gamma(\hat{S}_{\alpha k} B_{kj} + \tilde{w}_j/\tilde{w})} = \\
    &=\prod_{j=1}^5 \dfrac{\Gamma(\hat{S}_{\alpha k} B_{kj} + \tilde{w}_j/\tilde{w})}{\Gamma(1-n_a S_{ak}B_{kj} - \tilde{w}_j/\tilde{w})\gamma(\hat{S}_{\alpha k} B_{kj} + \tilde{w}_j/\tilde{w})},
\end{aligned}
\end{equation}
in the third equality we used that
\begin{equation}
\sum_{a=1}^h n_a - |\alpha| = \sum_{j=1}^5(n_a S_{ak} - \hat{S}_{\alpha k})B_{kj},\quad (n_a S_{ak} - \hat{S}_{\alpha k})B_{kj}\in \mathbb{Z}.
\end{equation}
Now we can rewrite (\ref{KalPotF}) as follows:
\begin{equation}
\begin{aligned}
e^{-K_C(X)} &= \sum_{\alpha=0}^{2h+1} \sum_{\substack{n_a,\bar{n}_a = 0\\{(n_a S_{ak} - \hat{S}_{\alpha k})B_{kj} \in \mathbb{Z}}\\{(\bar{n}_a S_{ak} - \hat{S}_{\alpha k})B_{kj} \in \mathbb{Z}}}}^\infty \Sigma_\alpha(\bar{n})\prod_{j=1}^5\dfrac{\gamma(\hat{S}_{\alpha k} B_{kj} + \tilde{w}_j/\tilde{w}) \Gamma(n_a S_{ak}B_{kj} + \tilde{w_j}/\tilde{w})}{\Gamma(\hat{S}_{\alpha k} B_{kj} + \tilde{w}_j/\tilde{w})}\prod_{a=1}^h \dfrac{(-1)^{n_a}\phi_a^{n_a}\bar{\phi}_a^{\bar{n}_a}}{n_a!\bar{n}_a!} =\\
&= \sum_{\substack{n_a,\bar{n}_a =0\\ (n_a-\bar{n}_a)S_{ak}B_{kj} \in \mathbb{Z}}}^\infty\prod_{j=1}^5 \dfrac{\Gamma(n_a S_{ak}B_{kj} + \tilde{w_j}/\tilde{w})}{\Gamma(1-\bar{n}_a S_{ak}B_{kj} - \tilde{w_j}/\tilde{w})}\prod_{a=1}^h \dfrac{(-1)^{n_a}\phi_a^{n_a}\bar{\phi}_a^{\bar{n}_a}}{n_a!\bar{n}_a!}.\label{KPotParF}
\end{aligned}
\end{equation}

The extension of the summation region can be explained as follows. Consider the particular term with some $n$ and $\bar{n}$ and assume that it is nonzero, for simplicity we denote
\begin{equation}
    \mathfrak{m}_j(n) = n_a S_{ak}B_{kj} + \tilde{w}_j/\tilde{w}.
\end{equation}
In the region of summation $\mathfrak{m}_j(n)$ and $\mathfrak{m}_j(\bar{n})$ can be integer only simultaneously, let $l$ be the number of integer elements of $\{\mathfrak{m}_j(n)\}$. So, due to the property \ref{A} there are at most $l/2$ singular gamma functions in the numerator of the chosen term of (\ref{KPotParF}), and at least $l/2$ of them in its denominator~--- the expression is always finite. As this term is nonzero, $l$ is even and exactly $l/2$ of integer elements of $\{\mathfrak{m}_j(n)\}$ and of $\{\mathfrak{m}_j(\bar{n})\}$ are positive. Hence as we have shown before $\mt{M}(nS) \neq 0$, $\mt{M}(\bar{n}S) \neq 0$ and therefore the corresponding monomials define non-trivial elements of $\mt{H}^5$. It means that all of the nonzero terms of (\ref{KPotParF}) are present in (\ref{KalPotF}) and the zero terms correspond to trivial elements of $\mt{H}^5$.

The main advantage of (\ref{KPotParF}) is that it does not contain terms with $\hat{S}$ and summation over $\alpha$, although it is not manifestly symmetric in $n_a$ and $\bar{n}_a$ unlike (\ref{KalPotF}). Besides, such a form will be much more convenient for comparison with the GLSM computation. 

Similar calculations can also be performed for the shifted $W_0$ (\ref{shift}). In order to find the cohomologic relations for $\prod_{j=1}^5 x_j^{b_j} d^5x$ we can use (\ref{cohomrel}) for $\prod_{j=1}^5 x_j^{b_j + i p_k M_{kj}}d^5x$ due to the pairing (\ref{pair}). It leads to the same shif $\tilde{w}_j/\tilde{w} \to \tilde{w}_j/\tilde{w} + i p_j$ as in the previous subsection. Besides, $\prod_{j=1}^5 x_j^{b_j} d^5 x$ is always nontrivial in case of nonzero $p_j$ (even if it cannot be reduced to $e_\alpha d^5x$ for some $\alpha$), so the basis $\{e_\alpha d^5x\}$ should be extended. The final result is as follows:
\begin{equation}
    e^{-K_C(X|p)} = \sum_{\substack{n_a,\bar{n}_a =0\\ (n_a-\bar{n}_a)S_{ak}B_{kj} \in \mathbb{Z}}}^\infty\prod_{j=1}^5 \dfrac{\Gamma(n_a S_{ak}B_{kj} + \tilde{w_j}/\tilde{w} + ip_j)}{\Gamma(1-\bar{n}_a S_{ak}B_{kj} - \tilde{w_j}/\tilde{w} - ip_j)}\prod_{a=1}^h \dfrac{(-1)^{n_a}\phi_a^{n_a}\bar{\phi}_a^{\bar{n}_a}}{n_a!\bar{n}_a!}.\label{shiftpot}
\end{equation}
We use $K_C(X|p)$ to denote this result simply for convenience: the zero locus (\ref{shift}) is not even well-defined in $\mathbb{P}^4_{w_1,\dots,w_5}$ if $p_j \neq 0$, so there is no connection with CY moduli spaces. However (\ref{shift}) can be viewed as a superpotential of some Landau-Ginzburg model, just as $W_0$.

\section{Mirror symmetry and GLSM partition function}\label{partfnc}
\subsection{Mirror GLSM}
The way to construct the mirror GLSM with $h$ of K\"ahler parameters was considered in \cite{Aleshkin:2019pmz,Aleshkin:2018tbx}. Let us clarify the connection between this approach and the one proposed by Batyrev \cite{Batyrev1993DualPA}. First, we need to represent $\mathbb{P}^4_{w_1,\dots,w_5}$ as a toric manifold defined by a fan. The one-dimensional cones of this fan are generated by vectors of the integral lattice $v_i \in \mathbb{N}^4$, $i=1,\dots,5$ such that $v_i w_i = 0$. These vectors should generate the lattice or its quotient by a finite group $G$ if we consider $\mathbb{P}^4_{w_1,\dots,w_5}/G$. All higher dimensional cones correspond to the faces of the simplex formed by $\{v_i\}$. One can also define an anticanonical polytope $\Theta_X=\{u:~\bl u,v_i \br \ge -1\}$, where $\bl,\br$ is a standard scalar product. The original construction of \cite{Batyrev1993DualPA} is based on the assumption that $\Theta_X$ is a lattice polytope~--- then the cones over faces of $\Theta_X$ define a mirror fan. The mirror family is represented by zeros of sections of the corresponding anticanonical bundle. Additionally, $\Theta_X$ should be reflexive. Both of these conditions are satisfied only if $w$ is divisible by $w_i$, i.e., when there is a Fermat polynomial $W_0$ \cite{Clader:2014kfa}. As we wish to consider more general cases, we need some generalization of this construction.

In \cite{ARTEBANI2016319} it was proposed to consider the convex hull of some of the integer points of $\Theta_X$ instead and construct a fan out of cones over its faces. In fact, such points correspond to the quasi-homogeneous monomials with respect to $\{w_i\}$. The exact map is as follows:
\begin{equation}
    n\mapsto \prod_{j=1}^5 x_j^{\bl n,v_j\br + 1},\quad n \in \mathbb{N}^4 \cap \Theta_X.
\end{equation}\label{degmap}
In particular, the image of the origin is $e_1$. If we choose the points corresponding to the monomials of $W_0$, we obtain precisely the fan for $\mathbb{P}_{\tilde{w}_1,\dots,\tilde{w}_5}^4$ (or its quotient by a finite group)~--- it is consistent with the BHK construction. However, we will also add the points corresponding to $e_a$ monomials for $a>1$ and subdivide the fan, assuming that \emph{all} of the chosen points generate one-dimensional cones, not just the ones being vertices of the hull. Hence the toric manifold we obtain is a blowup of one corresponding to the convex hull. We will see that such a method allows one to construct a GLSM with precisely $h$ K\"ahler parameters, which in some cases is a mirror to $X$.

The most suitable description of toric manifolds for our purposes is the one in terms of projective coordinates. In our case we have $h+4$ one-dimensional cones, hence there should be $h+4$ projective coordinates $z_1,\dots,z_5,z_7,\dots,z_h$. The reason for such peculiar numeration will become clear later. We also need to find the integral basis of linear relations between the vectors generating one-dimensional cones, the space of these relation $h$-dimensional. Due to (\ref{degmap}) we can use $M_{ij}-1$ and $S_{aj}-1$ instead of such vectors, so it is convenient to define the matrix $V_{\mu i}$, $\mu=1,\dots,h+5$ as follows:
\begin{equation}
    V_{ij} = M_{ij},\quad V_{5+a,j} = S_{aj}.
\end{equation}
We will encode the basis of relations into a matrix $Q_{\mu a}$:
\begin{equation}
    \sum_{\mu \neq 6} Q_{\mu a} (V_{\mu i} - 1) = 0.
\end{equation}\label{rel1}
As $V_{6 i} = S_{1 i} = 1$, it is equivalent to the basis of relations of $V_{\mu i}$:
\begin{equation}
    Q_{6a} = -\sum_{\mu \neq 6} Q_{\mu a},\quad Q_{\mu a}V_{\mu i} = 0.\label{Q6a}
\end{equation}
The toric manifold is defined as a quotient $(\mathbb{C}^{h+4} - Z)/\mathbb{C}^{*h}$ \cite{Hori:2003ic}, where the invariant set $Z$ is defined by the higher dimensional cones of the fan and $\mathbb{C}^{*h}$ acts as follows:
\begin{equation}
    z_\mu \mapsto \prod_{a=1}^h \lambda_a^{Q_{\mu a}} z_\mu,\quad \mu=1,\dots,5,7,\dots,h+5,~\lambda \in \mathbb{C}^{*h}.\label{cstar}
\end{equation}
We will denote it $\mt{Y}_{\text{B}}$. The section of its anticanonical bundle is a polynomial $\mt{W}(y)$ which has the following transformation properties under (\ref{cstar}):
\begin{equation}
    \mt{W} \mapsto \prod_{a=1}^h \lambda_a^{\sum_{\mu\neq 6} Q_{\mu a}} \mt{W}.\label{acan}
\end{equation}
So a representative of the mirror family is defined by $\mt{W}(y) = 0$. We can use (\ref{rel1}) to construct a particular example $\mt{W}_0$:
\begin{equation}
    \mt{W}_0(y) = \sum_{i=1}^5 \prod_{\mu \neq 6} z_\mu^{V_{\mu i}};
\end{equation}
In particular, in the chart $z_\mu \neq 0,~\mu>6$ (so we can set these $z_\mu$ to $1$ using the action of $\mathbb{C}^{*h}$) which is equivalent to the quotient of $\mathbb{P}^4_{\tilde{w}_1,\dots,\tilde{w}_5}$ by some finite group $G^T$, $\mt{W}_0(y)$ coincides with the transposed polynomial (\ref{trpol}). In \cite{Belavin:2020xhs} it was shown that $G^T$ is the same as in BHK construction.

Now we wish to construct a GLSM whose vacuum moduli space is also defined by $\mt{W} = 0$ in the same toric manifold. As we will see, in general it will be the case only for a particular chart. Toric manifolds arise naturally in case of GLSM \cite{Hori:2003ic}: the homogeneous coordinates correspond to the scalar fields, and the action of $\mathbb{C}^{*h}$ is defined by the charge matrix of $U(1)^h$ gauge group. The vacuum manifold however is given by a critical locus of an \emph{invariant} polynomial~--- a superpotential of the model. So we introduce a new coordinate $z_6$ and define
\begin{equation}
    \hat{\mt{W}}(y) = z_6 \mt{W},
\end{equation}
which is invariant with respect to the charge matrix $Q_{\mu a}$ due to (\ref{Q6a}) and (\ref{acan}). Its critical locus $Y$ is defined by the system $\mt{W} = 0$, $z_6 \p_\mu \mt{W} = 0$, $\mu \neq 6$. As $\sum_{\mu=1}^{h+5} Q_{\mu a} = 0$, this is indeed a CY manifold. 

Besides, the toric manifold $\mt{Y}_{\text{GLSM}}$ in this case is given in terms of symplectic quotient as a solution of the following equations:
\begin{equation}
\sum_{\mu=1}^{h+5} Q_{\mu a} |z_\mu|^2 = r_a,\quad r_a \in \mathbb{R},\label{momeq}
\end{equation}
considered up to $U(1)^h$ action:
\begin{equation}
 z_\mu \mapsto e^{2\pi i Q_{\mu a} \beta_a} z_\mu,\quad \beta_a \in \mathbb{R}.   
\end{equation}\label{U1}
Here $r_a$ are K\"ahler parameters, and their number coincides with the number of complex deformations considered in the previous section. Note that the corresponding fan is defined in $5$-dimensional space, as there are $h+5$ coordinates and $h$ relations. Hence, it is not a subdivision of the fan of $\mt{Y}_{\text{B}}$. The hypersurface $z_6 = 0$ defines the toric manifold with the same $1$-dimensional cones as $\mt{Y}_{\text{B}}$, as the relation matrices coincide, but the higher dimensional cones may differ.

In case of $\mt{Y}_{\text{B}}$ it is important to have integer $Q_{\mu a}$ in order to correctly define the action of $\mathbb{C}^{*h}$ (\ref{cstar}). On the other hand, the construction of $\mt{Y}_{\text{GLSM}}$ allows us to do an arbitrary real change of basis of relations, as (\ref{U1}) is still well-defined in this case. In fact, the most convenient expression for $Q_{\mu a}$ is generally non-integer:
\begin{equation}
    Q_{5+a,b} = -\delta_{a b},~ Q_{j a} = S_{ak} B_{kj}.\label{QExp}
\end{equation}
The equations (\ref{momeq}) are then as follows:
\begin{equation}
    \sum_{j=1}^5 S_{ak} B_{kj} |z_j|^2 - |z_{5+a}|^2 = r_a.
\end{equation}
Let us return to the chart $z_\mu = 1$, $\mu > 6$, considered previously for $\mt{Y}_B$. If we assume that $r_1 > 0$, at least one of $z_j$, $j = 1,\dots,5$ should be nonzero. Also note that $S_{1k}B_{kj} = \tilde{w}_j/\tilde{w}>0$ so the equation for $a=1$ has solutions. Hence, if in this case $\mt{W}$ is transverse with respect to $\{z_j\}$, then $\p_i \mt{W} \neq 0$. Therefore $z_6 = 0$~--- this hypersurface in the considered chart is once again precisely the $\mathbb{P}_{\tilde{w}_1,\dots,\tilde{w}_5}^4$ (or its quotient) from BHK construction, and the critical locus is defined by $\mt{W} = 0$. For simplicity, from now we will consider the following particular superpotential:
\begin{equation}
    \hat{\mt{W}}_0 = z_6 \mt{W}_0 = \sum_{i=1}^5\prod_{\mu=1}^{h+5} z_\mu^{V_{\mu i}}.
\end{equation}

The GLSM has multiple phases \cite{Witten:1993yc}, which are determined by values of $r_a$. In the previous section, we used the small complex structure decomposition ($\phi_a \to 0$) to find the K\"ahler potential. Hence, it is natural to expect that it should correspond to the partition function in the phase $r_a \ll 0$. Let us examine this phase in more detail.

First, if $S_{ak}B_{kj} \ge 0$, we have $z_{5+a} \neq 0~\forall a$. So we can fix $z_{5+a} = 1$ and what is left is the quotient of $\mathbb{C}^5$ by a finite group whose coordinates are $\{z_i\}$. Also, $\hat{\mt{W}}_0 = W_0^T$, so this phase corresponds to a Landau-Ginzburg model with superpotential $W_0^T$: the vacuum is given by critical points of $W_0^T$, and so it is $\{0\}$ due to transversality of $W_0^T$. It is consistent with the BHK construction, and in the next subsection, we will see that the partition function indeed coincides with (\ref{KPotParF}). This is always the case for Fermat polynomials, as $B_{jk} = \frac{1}{M_{jk}} \delta_{jk} > 0$, but in appendix \ref{A2} we show that there are less trivial examples as well. 

On the other hand, if there are $b$ and $j$ such that $S_{ak}B_{kj} < 0$, then $z_{5+b}$ might be zero, so we cannot fix all of $z_{5+a}$ anymore. Thus, we have a mixed phase. The partition function will have additional contributions concerning (\ref{KPotParF}), which can be attributed to different charts of $M_C[X]$.

\subsection{The partition function}
The partition function of GLSM on a sphere calculated in \cite{Doroud_2013, Benini_2014} is as follows:
\begin{equation}
    Z_{\text{GLSM}}(Y) = \sum_{Q_{\mu a} m_a \in \mathbb{Z}} \ili_{C_1}\dots \ili_{C_h}\dfrac{d\tau_1\dots d\tau_h}{(2\pi i)^h} e^{4\pi \tau_a r_a - i\theta_a m_a}\prod_{\mu = 1}^{h+5} \dfrac{\Gamma(Q_{\mu a}(\tau_a - m_a/2) + q_{\mu}/2)}{\Gamma(1-q_\mu/2 - Q_{\mu a}(\tau_a + m_a/2))}.\label{PartFnc1}
\end{equation}
Here the contours $\{C_a\}$ go upwards parallel to the imaginary axis. The parameters $q_\mu$ are R-charges of the scalars and should be chosen in such a way that the charge of superpotential is $2$. The convenient choice is as follows:
\begin{equation}
 q_i = 2\dfrac{\tilde{w}_i}{\tilde{w}},\quad q_{5+a} = 0,~a=1,\dots,h.\label{RC}
\end{equation} 
Using (\ref{QExp}), one can easily check that condition on total R-charge of $\hat{\mt{W}}$ is equivalent to its invariance with respect to $Q_{1\mu}$. In appendix \ref{B1} we show that different choice of $q_\mu$ leads to the same answer up to a product of holomorphic and antiholomorphic functions. The K\"ahler moduli space is complexified by an addition of theta-parameters $\theta_a$ to $r_a$. It is also convenient to introduce the holomorphic coordinates $\zeta_a$:
\begin{equation}
    \zeta_a = e^{-2\pi r_a + i\theta_a}.
\end{equation}
Also, in the original papers, this formula was derived assuming $Q_{\mu a} \in \mathbb{Z}$, in this case, the summation goes along the standard integral lattice. We used a generalization to an arbitrary basis of charges proposed in \cite{Aleshkin:2018tbx}.

Now we can rewrite (\ref{PartFnc1}) using (\ref{QExp}) and (\ref{RC}):
\begin{equation}
\begin{aligned}
    Z_{\text{GLSM}}(Y) &= \sum_{\substack{m_a\in\mathbb{Z}\\ m_a S_{ak}B_{kj}\in\mathbb{Z} }} \ili_{C_1}\dots \ili_{C_h}\dfrac{d\tau_1\dots d\tau_h}{(2\pi i)^h} \prod_{j = 1}^{5} \dfrac{\Gamma((\tau_a - m_a/2)S_{ak}B_{kj} +\tilde{w}_j/\tilde{w})}{\Gamma(1-\tilde{w}_j/\tilde{w}- (\tau_a + m_a/2)S_{ak}B_{kj})}\times\\
    &\times \prod_{a=1}^h \left[ \dfrac{\Gamma(-\tau_a + m_a/2)}{\Gamma(1+\tau_a + m_a/2)} \zeta_a^{-\tau_a + m_a/2} \bar{\zeta}_a^{-\tau_a-m_a/2} \right].
\end{aligned}\label{PartFnc2}
\end{equation}
As $r_a \ll 0$, we can close $C_a$ at $\text{Re}\, \tau_a \gg 0$, so the integral is equal to sum of residues in poles with $\text{Re}\, \tau_a \ge 0$ (the overall sign should be negative as the contour is oriented clockwise). First let us assume that $S_{ak}B_{kj} \ge 0$. Gamma functions have poles in integer non-positive points, so suppose that
\begin{equation}
 (\tau_a - m_a/2)S_{ak}B_{kj} +\tilde{w}_j/\tilde{w} = -n,~n\ge 0,~n\in \mathbb{Z}.
\end{equation}
Then we have:
\begin{equation}
1-\tilde{w}_j/\tilde{w}- (\tau_a + m_a/2)S_{ak}B_{kj} = 1+n-m_a S_{ak}B_{kj} = 1- 2\tau_a S_{ak} B_{kj}-2\tilde{w}_j/\tilde{w}. 
\end{equation}
Due to the $m_a$ summation region and as $\tau_a \ge 0$ we find that this expression is a non-positive integer. Hence, poles of numerators of  the first $5$ gamma functions are canceled by respective denominators. It means that we need to consider only poles of the following form:
\begin{equation}
-\tau_a + m_a/2 = -n_a,~n_a \ge 0,~n_a\in\mathbb{Z}.\label{inpoles}
\end{equation}
It is also convenient to change the summation variable. We define
\begin{equation}
    \bar{n}_a = \tau_a + m_a/2 = m_a + n_a.
\end{equation}
As $2\tau_a = n_a + \bar{n}_a$, all non-negative values of $\bar{n}_a$ satisfy $\tau_a \ge 0$. Besides, as $\bar{n}_a \in \mathbb{Z}$, terms with negative $\bar{n}_a$ have poles in denominator and hence do not contribute.

After calculating the residues and changing the summation variable to $\bar{n}_a$ we find:
\begin{equation}
    Z_{\text{GLSM}}(Y) = \sum_{\substack{n_a,\bar{n}_a =0\\ (n_a-\bar{n}_a)S_{ak}B_{kj} \in \mathbb{Z}}}^\infty \prod_{j=1}^5\dfrac{\Gamma(n_a S_{ak}B_{kj} + \tilde{w_j}/\tilde{w})}{\Gamma(1-\bar{n}_a S_{ak}B_{kj} - \tilde{w_j}/\tilde{w})} \prod_{a=1}^h \dfrac{(-1)^{n_a}\zeta_a^{-n_a}\bar{\zeta}_a^{-\bar{n}_a}}{n_a!\bar{n}_a!}.\label{ZGLSMres}
\end{equation}
We see that it coincides with the exponential of the K\"ahler potential (\ref{KPotParF}) if \begin{equation}
    \phi_a = \dfrac{1}{\zeta_a} \label{MirMap}.
\end{equation}
So the conjecture of \cite{J} is satisfied and the mirror map is defined by (\ref{MirMap}). Also, the initial expression (\ref{PartFnc2}) allows one to analytically continue this result to the region of large $\phi_a$ (or small $\zeta_a$) and thus explore the large complex structure regime by closing the contours at $\text{Re}\,\tau_a \ll 0$.

Now we move to the case with some of $S_{aj}B_{jk}$ being negative. Some poles of the first $5$ gamma functions in the numerator of (\ref{PartFnc2}) are no longer cancelled, so we have to take them into account as well. Still, the residue sum over the same poles (\ref{inpoles}) as before is once again the expression (\ref{ZGLSMres}), which is well-defined thanks to the property \ref{A} and coincides with (\ref{KPotParF}) if we use the mirror map. Also note that the modification of R-charges $q_j \to q_j + 2i p_j$ (which is formally not allowed) leads to (\ref{shiftpot}), so it is dual to the shift of $W_0$ (\ref{shift}).

Now we consider the generic situation when gamma function with numbers $\mu_a$ have poles. We also change the R-charges in such a way that $q_{\mu_a} = 0$. Let us denote $\hat{Q}_{ab} = Q_{\mu_a b}$, so the poles are defined by
\begin{equation}
    \hat{Q}_{ab}(\tau_b - m_b/2) = -n_a.\label{gpoles}
\end{equation}
Similarly to the previous computation we also define $\bar{n}_a = -\hat{Q}_{ab} (\tau_b + m_b/2)$. However, not all integral non-negative sets $\{n_a\}$ and $\{\bar{n}_a\}$ are possible, as
\begin{equation}
    n_a + \bar{n}_a = -2\hat{Q}_{ab} \tau_b
\end{equation}
and $\tau_b \ge 0$. It means that unless $\hat{Q}$ is non-degenerate and the elements of $\hat{Q}^{-1}$ are all non-positive, some of $\{n_a\}$ and $\{\bar{n}_a\}$ cannot be realized. If we define a \emph{negative cone} in $\mathbb{R}^h$ as the one spanned by $\{-e_a\}$, where $\{e_a\}$ is a standard basis, this condition can be reformulated. Namely, the elements of $\hat{Q}^{-1}$ are non-positive iff the cone spanned by rows of $\hat{Q}$ contains the negative cone. In fact, if it is not satisfied, one can show that the corresponding residue sum vanishes \cite{Aleshkin:2022}. To demonstrate this property we provide a simple example in appendix \ref{VPoles}.

Having $\hat{Q}^{-1}_{ab} \le 0$ we can do the following change of coordinates:
\begin{equation}
\tau_a = -\hat{Q}_{ab}^{-1}\tilde{\tau}_b,\quad m_a = -\hat{Q}_{ab}^{-1} \tilde{m}_b,\quad \zeta_a = \prod_{b=1}^h \tilde{\zeta}_{b}^{-\hat{Q}_{ba}}\label{CoordCh}
\end{equation}
which does not change the way we close the contours $C_a$. The result of this change coincides with formula (\ref{PartFnc1}) written in a different basis of charges, where $Q_{\mu_a b} = -\delta_{ab}$. It has an interesting interpretation in terms of polynomial $W$. Namely, such choice of charges (or relations) means that we consider $\prod_{j=1}^5 x_j^{V_{\mu_a j}}$ as deformations and the sum of the rest of the monomials as the reference polynomial $\tilde{W}_0$. So we define $\tilde{S}_{aj} = V_{\mu_a j}$, compose a matrix $\tilde{M}_{ij}$ from the rest of the rows of $V$ and consider
\begin{equation}
    \tilde{W}_0 = \sum_{i=1}^5 \prod_{j=1}^5 x_j^{\tilde{M}_{ij}},\quad \tilde{W} = \sum_{i=1}^5\prod_{j=1}^5 x_j^{\tilde{M}_{ij}} + \sum_{a=1}^h \tilde{\phi}_a \prod_{j=1}^5 x_j^{\tilde{S}_{aj}}.
\end{equation}
It defines a different chart in the complex moduli space with the reference point $\tilde{W}_0 = 0$. This polynomial can be obtained from $W$ by coordinate rescaling $x_j \to \prod_{a=1}^h \phi_a^{R_{ja}}x_j$ such that the coefficients at $\prod_{j=1}^5 x_j^{\tilde{M}_{ij}}$ become unity. Then $\tilde{\phi}_a$ can be expressed in terms of $\phi_a$ and as we show in appendix \ref{B3} these expressions are consistent with the mirror map (\ref{MirMap}) and the change (\ref{CoordCh}):
\begin{equation}
    \tilde{\phi}_a = \prod_{b=1}^h \phi_b^{-\hat{Q}_{ba}^{-1}} = \dfrac{1}{\tilde{\zeta}_a}.\label{MirMap2}
\end{equation}
Thus, it is natural to assume that the GLSM we constructed is mirror to a collection of charts of the complex moduli space $M_C(X)$: each chart corresponds to a particular set of poles (\ref{gpoles}). The charts themselves are defined by the condition $\hat{Q}^{-1}_{ab} \le 0$.

Note that all degrees in (\ref{MirMap}) here are non-negative, so $\tilde{\phi} \to 0$ when $\phi \to 0$. Besides, one can define the R-charges as $\tilde{q}_j = 2 \sum_{i=1}^h \tilde{B}_{ij}$ (we changed the order of gamma function in the same way we changed the rows of $V_{\mu i}$), where $\tilde{B} = \tilde{M}^{-1}$, and obtain the formula similar to (\ref{ZGLSMres}) for a given set of poles (\ref{gpoles}). If we also shift $\tilde{q}_j \to \tilde{q}_j + 2 i p_j$, the residue sum we consider becomes as follows:
\begin{equation}
Z_{\text{GLSM}}(Y|\{\mu_a\};p) = \sum_{\substack{n_a,\bar{n}_a =0\\ (n_a-\bar{n}_a)\tilde{S}_{ak}\tilde{B}_{kj} \in \mathbb{Z}}}^\infty \prod_{j=1}^5\dfrac{\Gamma(n_a \tilde{S}_{ak}\tilde{B}_{kj} + \tilde{q}_j + ip_j)}{\Gamma(1-\bar{n}_a \tilde{S}_{ak}\tilde{B}_{kj} - \tilde{q}_j - ip_j)} \prod_{a=1}^h \dfrac{(-1)^{n_a}\tilde{\zeta}_a^{-n_a}\bar{\tilde{\zeta}}_a^{-\bar{n}_a}}{n_a!\bar{n}_a!}.\label{PartFuncP}
\end{equation}
We see that it is indeed dual to the result of the geometric computation for $\tilde{W}$ given by (\ref{shiftpot}) with appropriate change of parameters. The duality is established by the mirror map (\ref{MirMap2}). However, we are also interested in the limit $p_j \to 0$ in order to restore the connection with $M_C[X]$. As $\tilde{W}_0$ is not generally transverse, the property \ref{A} no longer holds. It was essential for existence of such a limit, so without it some terms of (\ref{PartFuncP}) might be singular when $p_j = 0$. They correspond to non-simple poles of the integrand of (\ref{PartFnc1})~--- the residue sum at $p_j = 0$ should be computed more accurately. For instance, non-simple poles lead to terms logarithmic in $\tilde{\zeta}_a$ and hence logarithmic in $\tilde{\phi}_a$. It means that the behaviour of the residue sum at $\tilde{\phi}_a \to 0$ is singular. This is not surprising: if $\tilde{W}_0$ is not transverse, it corresponds to the singular point of the moduli space. So it is natural to assume that the residue sum represents the actual K\"ahler potential computed in the corresponding chart of $M_C[X]$.

Another interesting question is whether it is possible to construct a GLSM which is mirror just to one initial chart of $M_C[X]$ for a generic transverse $W_0$. It seems that the answer is no, but there is a formal procedure which leads to the desired result. Namely, replace $S_{ak}B_{kj}$ in the arguments of gamma functions in (\ref{PartFnc2}) with some generic $Q_{ja} \ge 0$ (the summation region stays the same). Then only the set of poles (\ref{inpoles}) has a nonzero contribution, and the complete residue sum is just slightly modified (\ref{ZGLSMres}). By subsequent analytic continuation of $Q_{ja}$ back to $S_{ak} B_{kj}$ we restore (\ref{ZGLSMres}) and hence obtain $e^{-K_C(X)}$ calculated in the appropriate chart. In principle, this procedure can be used to study the large complex structure regime as well: the analytic continuation from non-negative $Q_{ja}$ should be done \emph{after} the calculation of the residue sum as before.

\section{Conclusion}
We discuss the method of computation of K\"ahler potential on complex structure moduli space for CY-manifolds in weighted projective spaces. The particular case of transverse polynomials is considered. We compute the periods and the real structure structure matrix in a specific basis of cycles; the general formula for the potential is derived. We also show that the obtained expressions are well-defined thanks to the specific property of transverse polynomials. Besides, a particular logarithmic deformation of the original polynomial $W_0$ is discussed to avoid the situations when the chosen basis of real cycles is ill-defined.

Then we construct the mirror GLSM using Batyrev's approach to mirror symmetry. We show that in the case of non-negative charges $Q_{ja}$, it has a Landau-Ginzburg phase consistent with the Berglund-Hubsch-Krawitz mirror construction. We demonstrate that the GLSM partition function coincides with the exponential of the K\"ahler potential and allows one to define a mirror map.

Finally, we study the case of some of the charges $Q_{ja}$ being negative. We argue that it is a mirror to a collection of charts of the complex moduli and show that such an interpretation is consistent with the obtained mirror map. The deformation of R-charges is demonstrated to be dual to the discussed logarithmic deformation of $W_0$. Finally, a way to obtain a mirror to only one chart via analytical continuation from the model with positive charges is provided. We speculate on the possibility to use this method to investigate the large complex structure regime as well.

\section{Acknowledgements}
We would like to thank A. Belavin and K. Aleshkin for valuable discussions. We are also grateful to A. Artemev for careful reading of the paper. This work was supported by the Russian Science Foundation grant (project no. 18-12-00439).
\begin{appendices}
\section{An important property of transverse polynomials}\label{A}
We wish to prove the following property of the transverse polynomial $W_0$: \emph{If $\{n_j\}$ is a set of non-negative integers, then the number of positive integer elements of $\{n_k B_{kj} + \tilde{w}_j/\tilde{w}\}$ is always not less then the number of negative ones}. For simplicity we denote $\Omega_j = n_k B_{kj} + \tilde{w}_j/\tilde{w}$. According to \cite{krawitz2009fjrw}, transverse polynomials consist of the following elementary blocks:
\begin{equation}
\begin{aligned}
x_1^{a_1} + \dots + x_K^{a_K} &- \text{Fermat}\\
x_1^{a_1}x_2+x_2^{a_2}x_3+\dots+x_K^{a_K}&- \text{chain}\\
x_1^{a_1}x_2+x_2^{a_2}x_3+\dots+x_K^{a_K}x_1&- \text{loop},
\end{aligned}
\end{equation}
here $K \le 5$. The matrices $M$ and $B$ are block-diagonal, so we can consider these blocks separately. Hence, the limits of all summations over repeated indices in this sections are $1$ and $K$. We will also slightly abuse the notation, denoting the particular block by $W_0$. The property under consideration is obviously satisfied for Fermat blocks, as $B_{kj} = \frac{\delta_{kj}}{a_k} \ge 0$. The cases of other two types are less trivial.

To proceed, we will need an auxiliary observation. Namely, note that for a given $l$ the only nontrivial elements of the $l$-th column of $M$ are $M_{l-1,l}$ and $M_{ll}$ (we identify $l=-1$ with $l=K$). As
\begin{equation}
\Omega_j M_{jl} = n_l + 1 > 0,
\end{equation}
the following condition is satisfied:
\begin{equation}
    \Omega_l > 0~\text{or}~\Omega_{l-1} > 0.\label{auxprop}
\end{equation}
It can be improved in case $l=1$ for chain polynomials, as for them $B_{j1} = \frac{\delta_{j1}}{a_1} \ge 0$:
\begin{equation}
    \Omega_1 > 0.\label{auxpropchain}
\end{equation}

Another important fact is that $x_j \mapsto e^{2\pi i B_{kj}} x_j$ is a symmetry of the transposed polynomial $W_0^T$. In particular, it means that $x_j \mapsto e^{2\pi i \Omega_j} x_j$ is also its symmetry due to the definition of $\tilde{w}_j/\tilde{w}$ (\ref{tilw}). We denote this symmetry generator by $g(\Omega)$. The transposed chain and loop polynomials are as follows:
\begin{equation}
 W_{0,\text{chain}}^T = x_1^{a_1} + x_1 x_2^{a_2} + \dots + x_{K-1} x_K^{a_K},\quad W_{0,\text{loop}}^T = x_K x_1^{a_1} + x_1 x_2^{a_2} + \dots + x_{K-1} x_K^{a_K}. \label{transpChainLoop}
\end{equation}
Now suppose that $\Omega_m \in \mathbb{Z}$, then $g(\Omega)$ acts trivially on $x_m$. As $g(\Omega)$ is a symmetry of $W_0$, from (\ref{transpChainLoop}) we inductively find that $g(\Omega)$ acts trivially on $x_j$ for $1 \le j \le K$ if $W_0$ is a loop and for $1 \le j \le m$ if $W_0$ is a chain (in this case we assume that the action on $x_{m+1}$ is nontrivial). Hence, $\Omega_j \in \mathbb{Z}$ for corresponding $j$. To conclude the proof, assume the opposite to the property \ref{A}:
\begin{itemize}
    \item For loop polynomials it means that there are more than $K/2$ of non-positive $\Omega_j$, $1 \le j\le K$. Hence, there should be at least two consecutive non-positive elements of $\{\Omega_j\}$ ($\Omega_1$ and $\Omega_K$ are assumed to be consecutive as well). It is impossible due to (\ref{auxprop}).
    \item For chain polynomials it means that there are more that $m/2$ of non-positive $\Omega_j$, $2 \le j \le m$, here we used (\ref{auxpropchain}) to exclude $\Omega_1$. Again, it means that there should be at least two consecutive non-positive elements of $\{\Omega_j\}$, $2 \le j \le m$, which is forbidden by (\ref{auxprop}).
\end{itemize}
In both cases we obtained a contradiction, thus the proof is finished.

\section{Properties of partition function and relevant examples}
\subsection{Non-fermat polynomial with non-negative charges $Q_{ja}$}\label{A2}
Consider the non-Fermat transverse polynomial
\begin{equation}
W_0 = x_1^4 + x_2^4x_1 + x_3^5+x_4^5+x_5^5
\end{equation}
in $\mathbb{P}_{3,5,4,4,4}$ with an additional $\mathbb{Z}_5$ quotient defined by the following generator:
\begin{equation}
g: \begin{pmatrix}
    x_1\\
    x_2\\
    x_3\\
    x_4\\
    x_5
\end{pmatrix} \mapsto \begin{pmatrix}
    x_1\\
    \w x_2\\
    x_3\\
    x_4\\
    \w^{-1} x_5
\end{pmatrix},\quad \w = e^{2\pi i/5}.
\end{equation}
One can find that $h=5$ and the deformation and charge matrices are as follows:
\begin{equation}
S = \begin{pmatrix}
1&1&1&1&1\\
0&0&2&3&0\\
0&0&3&2&0\\
1&1&0&2&1\\
1&1&2&0&1\\
\end{pmatrix},\quad SB = \dfrac{1}{5}\begin{pmatrix}
1&1&1&1&1\\
0&0&2&3&0\\
0&0&3&2&0\\
1&1&0&2&1\\
1&1&2&0&1\\
\end{pmatrix},
\end{equation}
so the charges $Q_{ja} = S_{ak} B_{kj}$ are all non-negative.

\subsection{The dependence on R-charges}\label{B1}
Let us consider the arbitrary choice of R-charges $\hat{q}_\mu$. They should satisfy $\hat{q}_\mu V_{\mu i} = 2$ and we already know a particular solution (\ref{RC}) of this equation and a basis of relations for $V_{\mu i}$. Hence,
\begin{equation}
    \hat{q}_\mu = q_\mu + 2Q_{\mu a} \chi_a,
\end{equation}
where $q_\mu$ are defined by (\ref{RC}). Looking at (\ref{PartFnc1}), we see that we can get rid of $Q_{\mu a} \chi_a$ by change $\tau_a \to \tau_a - \chi_a$ (we assume that the contours $C_a$ do not cross poles after this change). So the partition function simply acquires a factor $\prod_{a=1}^h (\zeta_a \bar{\zeta}_a)^{\chi_a}$ which does not change the K\"ahler potential.

\subsection{An example of vanishing poles}\label{VPoles}
Consider the following integral:
\begin{equation}
    I = \ili_{C_1} \ili_{C_2} \dfrac{d\tau_1 d\tau_2}{(2\pi i)^2} \Gamma(-\tau_1) \Gamma(-\tau_1-\tau_2) e^{-r_1\tau_1-r_2\tau_2},\quad r_a > 0.
\end{equation}
The cone spanned by vectors $(-1,0)$ and $(-1,-1)$ clearly does not contain the negative cone. The contours go parallel to the imaginary axis slightly left to $\text{Re} \tau_a = 0$. We close $C_a$ at $\text{Re}\tau_a \gg 0$ and calculate the $\tau_1$ integral first:
\begin{equation}
I = \ili_{C_2} \dfrac{d\tau_2}{2\pi i} \sum_{n\ge 0} (-1)^n \dfrac{\Gamma(-n-\tau_2)}{n!} e^{-n r_1-\tau_2 r_2} + \ili_{C_2} \dfrac{d\tau_2}{2\pi i} \sum_{k\ge 0} (-1)^k \dfrac{\Gamma(\tau_2-k)}{k!}e^{-\tau_2 r_2} e^{\tau_2 r_1 - k r_1}.
\end{equation}
We numerate the poles as follows: $\tau_2 = m$ and in the case of the second integral $n = k-m\ge 0$. So we find:
\begin{equation}
I = \sum_{n,m=0}^{\infty} \dfrac{(-1)^n(-1)^{n+m}}{n!(n+m)!} e^{-nr_1-mr_2} - \sum_{n,m=0}^{\infty} \dfrac{(-1)^n(-1)^{n+m}}{n!(n+m)!} e^{-nr_1-mr_2} = 0,
\end{equation}
as we expected.

\subsection{Change of variables}\label{B3}
We will proof the consistency with mirror map inductively. Namely, consider the case
\begin{equation}
    \mu_a = \{6,7,\dots,h-1,1\},
\end{equation}
i.e. when we exchange the first and last rows in $V_{\mu i}$. The rescaling $x_j \to  \prod_{a=1}^h \phi_a^{R_{ja}}x_j$ should preserve the coefficients at $\prod_{j=1}^5 x_j^{M_{ij}}$ for $i>1$ and transform the coefficient $\phi_h$ at $\prod_{j=1}^5 x_j^{S_{hj}}$ to unity. The matrix $R_{ja}$ satisfying this requirements is as follows:
\begin{equation}
    R_{ja} = B_{jk}U_{ka},\quad U_{ja} = -\dfrac{\delta_{j,1} \delta_{a,h}}{S_{hk}B_{k1}}
\end{equation}
Using this expression one can easily find $\tilde{\phi}_a$: for $a < h$ they are coefficients at the same monomials as $\phi_a$ and $\tilde{\phi}_h$ is a coefficient at $\prod_{j=1}^5 x_j^{M_{1j}}$. We find:
\begin{equation}
    \tilde{\phi}_a = \phi_a \prod_{b=1}^h \phi_b^{S_{aj}R_{jb}} = \phi_a \phi_h^{-S_{ak}B_{k1}/S_{hl}B_{l1}},~a<h,\quad \tilde{\phi}_h = \prod_{a=1}^h \phi_a^{M_{1j}R_{ja}} = \phi_h^{-1/S_{hk}B_{k1}}.
\end{equation}
The matrix $\hat{Q}_{ab}$ and its inverse are as follows:
\begin{equation}
    \hat{Q}_{ab} = \begin{cases}
    -\delta_{ab},&a<h\\
    S_{bk}B_{k1},&a=h
    \end{cases},\quad \hat{Q}^{-1} = \begin{cases}
    -\delta_{ab},&a<h\\
    S_{bk}B_{k1}/S_{hk}B_{k1},&a=h
    \end{cases}.
\end{equation}
From these expressions one can see that (\ref{MirMap}) is satisfied. The case of a more general change can now be proven inductively.

\end{appendices}
\printbibliography
\end{document}